# Human-Generative AI Collaborative Problem Solving Who Leads and How Students Perceive the Interactions

Gaoxia Zhu
*National Institute of Education*
Nanyang Technological University
Singapore
gaoxia.zhu@nie.edu.sg

Vidya Sudarshan
College of Computing and Data Science
Nanyang Technological University
Singapore
vidya.sudarshan@ntu.edu.sg

Jason Fok Kow
College of Computing and Data Science
Nanyang Technological University
Singapore
jason.fokkow@ntu.edu.sg

Yew Soon Ong*
College of Computing and Data Science
Nanyang Technological University
Singapore
asysong@ntu.edu.sg

*Abstract*—This research investigates distinct human-generative AI collaboration types and students' interaction experiences when collaborating with generative AI (i.e., ChatGPT) for problem-solving tasks and how these factors relate to students' sense of agency and perceived collaborative problem solving. By analyzing the surveys and reflections of 79 undergraduate students, we identified three human-generative AI collaboration types: even contribution, human leads, and AI leads. Notably, our study shows that 77.21% of students perceived they led or had even contributed to collaborative problem-solving when collaborating with ChatGPT. On the other hand, 15.19% of the human participants indicated that the collaborations were led by ChatGPT, indicating a potential tendency for students to rely on ChatGPT. Furthermore, 67.09% of students perceived their interaction experiences with ChatGPT to be positive or mixed. We also found a positive correlation between positive interaction experience and a sense of positive agency. The results of this study contribute to our understanding of the collaboration between students and generative AI and highlight the need to study further why some students let ChatGPT lead collaborative problem-solving and how to enhance their interaction experience through curriculum and technology design.

*Keywords—Human-generative AI collaboration, ChatGPT, problem-solving, agency, overreliance, higher education*

## I. INTRODUCTION

In an era where the interplay between humans and artificial intelligence (AI) is increasingly prominent, understanding human-AI collaboration types and relevant impacts is crucial. AI, particularly generative AI and large language models like the popular ChatGPT, excels in repetitive, data-intensive tasks, while humans bring creativity, empathy, and common sense [1] [2]. Together, humans and AI are envisioned to potentially form synergistic teams capable of enhanced collaborative problem-solving. In this paper, we present a study on when undergraduate students collaborate with ChatGPT for problem solving, who leads the interaction, how students perceive the interaction, and the implications of this augmentation from AI on students' agency and collaborative problem-solving.

A challenging area for human-AI collaboration is solving complex problems [3]. Collaborative problem-solving, a critical 21st-century skill, involves groups working together to find solutions to complex problems [4]. Previously, the collaboration tended to be only among humans with shared interests or goals. In the face of increasingly capable AI on complex tasks, AI-enabled collaboration and collaboration between humans and AI draw increasing attention and research interests [5]. AI can serve as a collaborator in problem solving [6]. Meanwhile, it is critical to understand how students collaborate with these technologies and the relevant impacts on students' agency to ensure they empower rather than overshadow students' agency [7]. Agency involves making informed decisions, initiating actions, and controlling one's learning journey, a vital aspect of student learning [3] and deserves more attention when students collaborate with powerful generative AI.

In the subsequent sections, the results of our study on how students perceive their collaboration with ChatGPT regarding contributions and interaction experiences and how these perceptions relate to their agency and collaborative problem-solving shall be presented. It is hoped that this study will provide insights into conducive collaboration types between human and generative AI and inspire further research on human-generative AI collaboration and its impacts on students' learning, problem-solving, and psychological factors such as agency.

## II. RELATED WORK

### A. Human-AI Collaboration

Human-AI collaboration is about humans and AI working together on tasks under the underlying reasoning that they complement one another regarding strengths and weaknesses [8] [9]. AI, especially generative AI like ChatGPT, is good at repetitive and data-driven tasks, has strength in language comprehension and reasoning, and can engage in human-like conversations, whereas humans have common sense and are creative and empathic but may not be as good at storing and processing information [1] [2]. Therefore, they may work as teammates and engage in collaborative problem-solving.

Previous studies suggest different human-AI collaboration types or patterns in the education field. Based on a literature review, Ouyang and Jiao [10] summarized three paradigms of AI in education: 1) AI-directed, learner-as-recipient, 2) AI-supported, learner-as-collaborator, and 3) AI-empowered, learner-as-leader. They indicated that current development is moving towards Paradigm 3, emphasizing learners' agency, student-centered learning, empowerment, and reflections. Liu and Bridgeman [11] proposed four unique types of human-generative AI interactions: significant contributions from both AI and humans ("create it with me"), predominant AI input with minimal human involvement ("create it for me"), major human input with minor AI assistance ("help me brainstorm, plan, and polish"), and minimal input from both AI and humans. Cheng and Zhang [6] interviewed 63 university students after they engaged in ChatGPT-enabled group discussions. They found that ChatGPT might function as a tool, a teammate, or a form of superintelligence during collaboration, as determined through qualitative analysis of the interview data. However, Ouyang and Jiao's study [10] was based on a literature review, and Liu and Bridgeman's proposal [11] was a theoretical conjecture. Cheng and Zhang [6] did not instruct participants on the purpose of using the ChatGPT-enabled collaboration system but asked them to complete group discussions with its support. The various patterns of human-generative AI collaboration in the problem-solving context and their impacts warrant further investigation.

Human-AI collaboration has some benefits, such as saving human labor and avoiding humans' blind spots [12]; however, some challenges also exist. For instance, students may rely on AI assistance rather than learning from it; AI may limit the development of students' agency and essential skills for the future, decrease unique human knowledge, make humans act like machines or cyborgs, and raise ethics and authenticity issues [7] [13] [14]. In education, attention should be paid to using AI to assist and empower learners without sacrificing their leadership and agency. However, how AI and humans collaborate to solve problems and the impacts of the collaboration types remain unclear and need further research [8] [15].

*B. Sense of Agency*

Agency is about students' capacity to make responsible decisions, initiate their actions, and control their learning journeys [16] [17]. It includes indicators of an agent having intentions and purposes in pursuing ideas, being motivated and willing to make choices, persisting and acting on intentions, and interacting and negotiating in the learning context [17]. Having a sense of agency is closely related to student's ability to control their behavior, make informed decisions, and maneuver through complex social environments [18], which may be even more critical when generative AI like ChatGPT becomes increasingly powerful and capable of taking over some tasks previously can only be done by humans (e.g., translation and creative writing). Therefore, it is essential to study the impacts of AI on students' agency, with the goal of designing technological and pedagogical support for meaningful learning and enhanced agency.

*C. Collaborative Problem Solving*

Collaborative problem-solving is an essential 21st-century skill for students to survive and thrive in this rapidly developing society and workplace [4]. It is the process of a group of people attempting to find appropriate solutions to a given problem or group goal through systematic observation and critical thinking [19] [4]. Collaborative problem-solving involves students working in groups to collaboratively understand, integrate their contributions, and resolve a problem whose solution is not apparent and cannot be solved by single individuals [20].

Problem solving is a major challenging area for the hybrid team of humans and AI [3]. Memmert and Bittner [3] conducted a literature review on human-AI collaboration for complex problem solving and only identified 22 publications. They suggested identifying suitable research contexts (e.g., "tasks that so far seem to be at the core of human intellect" [2]) and considering the complexity of tasks to study the characteristics and patterns of human-AI collaboration. Therefore, this study selected a maze design and solving task which involves students' skills such as computational thinking, design, visual programming, and decision making, which are core human intellect. This exploratory study moves one step forward by studying the collaboration patterns between students and generative AI, which is more versatile than AI tools designed for specific tasks and purposes.

*D. Research Questions*

To summarize, related work suggests the existence of different types of human-AI collaboration patterns and the importance of students' agency and collaborative problem-solving skills. However, it remains unclear how students think of their collaboration with ChatGPT regarding contribution and interaction experiences and how these variables relate to their agency and collaborative problem-solving. This study aims to address these research gaps by answering the two following questions:

RQ1: Do distinct human-generative AI collaboration types arise when undergraduate students use ChatGPT for collaborative problem-solving?

RQ2: What are undergraduate students' interaction experiences with ChatGPT for collaborative problem-solving?

RQ3: Whether human-generative AI collaboration types, students' interaction experiences and sense of agency predict their collaborative problem-solving?

## III. METHODS

*A. Participants*

The participants of this study were 79 (44 female) undergraduate students taking an interdisciplinary learning course on digital literacy at an autonomous university in Singapore. The course is open to year one or two undergraduate students to cultivate their problem-solving, decision-making, and interdisciplinary collaboration skills with responsible, ethical, and legal use of digital technologies and tools. The participants came from various colleges across the university: 12 were from Business, 9 from Humanities, Arts and Social Sciences, 33 from Engineering, including Computer Science, and 25 from Science. Among 904 students who completed an individual survey regarding their reflections on using ChatGPT for collaborative maze design and solving, 79 gave us consent to analyze the data after receiving a recruitment email, making the study's final sample. This study was approved by the IRB of the authors' institution.

*B. Procedures*

In a two-hour session class, the participants first engaged in an activity on applying computational thinking skills to design and solve a maze in 3D Maze with ChatGPT in small groups. The groups were usually made up by four to six students from different schools and with various disciplinary backgrounds. 3D Maze is a gamified environment where students can design and solve mazes (see Fig. 1).

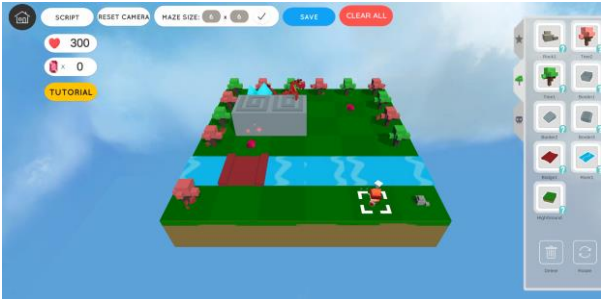

Fig. 1. The interface of 3D Maze

ChatGPT is a large language model developed by OpenAI. It can generate human-like text based on input prompts, answer questions, and perform natural language processing tasks. It was made clear to the students that they could decide when to use ChatGPT, for what purposes and how. Doing so allows us to study how students applied ChatGPT in problem-solving tasks in the natural setting as they desired to, without being instructed to follow specific application guidelines.

The following were the task requirements:

1. The maze's width and length should be at least 20 units.
2. The maze should contain all environmental assets (e.g., green tree, cherry blossom, rock, block A, block B, river, bridge)
3. The maze should be challenging and engaging.
4. Optional: the maze can contain two gems (gems need to be collectible in maze solving)
5. After finishing the maze creation in 3D Maze, each group needs to solve their own maze.

Toward the end of the activity, students completed an individual survey on collaborative problem-solving and sense of agency. They also responded to two open-ended questions on their own and ChatGPT's contribution to the maze design and solving task, as well as their interaction experiences with ChatGPT during the process.

*C. Data collection*

The data sources of this study are: (1) students' responses to the Collaborative Problem-Solving and Sense of Agency surveys and (2) their written responses to two reflection questions, which are "How would you describe your interaction with ChatGPT during the maze design task? In terms of contributions, did you feel that you, ChatGPT, or both equally contributed to the design of the maze? Please elaborate" and "What aspects of your interaction with ChatGPT today did you enjoy or find lacking?"

The Collaborative Problem-Solving scale was adapted from the TEAM Collaboration Questionnaire: Integration subscale [20]. Students were asked to respond to five items on a 5-point Likert Scale, 1 representing strongly disagree and 5 denoting strongly agree. Some example questions are "The solutions that my team and I found satisfied our needs" and "I applied knowledge and ideas from my subject field(s) when I approached the activities." Based on participants' responses, the Cronbach's alpha for this factor is 0.85, indicating the reliability of this measurement [21].

The Sense of Agency scale was slightly modified based on the sense of agency scale developed by Tapal et al. [22]. The original scale is a validated survey comprised of two factors: Sense of Positive Agency and Sense of Negative Agency. Each factor was measured using six items on a 5-point Likert Scale (1 denotes strongly disagree, whereas 5 represents strongly agree). An example question for measuring Sense of Positive Agency is "My behavior was planned by me from the very beginning to the very end," and an item for measuring Sense of Negative Agency is "I was just an instrument in the hands of ChatGPT." Cronbach's alpha values for the Sense of Positive Agency and Sense of Negative Agency were 0.90 and 0.85, respectively, suggesting the reliability of the two sub-scales.

*D. Data analysis*

To respond to RQ1 concerning human-generative AI collaboration types, we conducted a content analysis of students' written reflections on their own and ChatGPT's contributions to the maze design and solving task. As shown in Table 1, driven by the data and informed by the literature on human-AI collaboration [11] [6], we developed a coding framework that includes three human-generative AI collaboration types: even contribution, human leads, and AI leads. Based on the data, an unsure category in which students did not clearly describe the proportion of their and ChatGPT's contributions emerged. Two researchers independently coded each student's reflection and reached an agreement of 83.54%. Then, they resolved all the differences through discussions.

Similarly, to examine RQ2 regarding students' interaction experiences with ChatGPT, we analyzed students' reflections on their interaction experiences with ChatGPT using the coding framework shown in Table 2. Table 2 was driven by the data and the consideration of AI's positive and negative impacts on learning [13] [14]. The students mainly had positive, mixed, and negative interaction experiences with ChatGPT. There were also some cases where it was difficult to understand their sentiment based on the written reflections, or their writing was too brief, coded as "unsure." Two researchers coded the data separately and reached an agreement of 84.81%. They then discussed all the differences to reach an agreement.

TABLE 1. CODING FRAMEWORK OF HUMAN-GENERATIVE AI COLLABORATION TYPES

| Human-generative AI collaboration types | Descriptions | Examples |
| --- | --- | --- |
| **Even contribution** | Both students and ChatGPT contribute to the task equally. We equally value decision-making and repetitive work (heavy lifting) if students mention this part. They may use keywords such as equal, balanced, and partnership. | "I think both me and GPT equally contributed to the design of the maze, because I do the instructing and GPT did all the labor work." <br> "It was interesting because GPT's answer always managed to amaze me. However, I think we both contributed equally as I did all the decision-making on whether I would use GPT's suggestions or not, which means I also came up with my own ideas." |
| **Human leads** | Students contribute more, while ChatGPT contributes less. We can see students' efforts in applying critical thinking and evaluating ChatGPT's results. They may use keywords such as assistant, instruct, follow, f, modify prompts, supplement, supervise, guide, contribution was minimal. | "ChatGPT is not really helpful in designing the maze. The answers that GPT gave me were not satisfying. My group and I decided to design the maze by ourselves." <br> "I felt that I contributed more than GPT because GPT did not interpret my command correctly, and eventually all the maze ideas were purely mine. GPT creates not even a single idea in my experience regarding the maze design." |
| **AI leads** | ChatGPT contributes more to the task than students and does the task for students instead. (e.g., I hardly contribute) | "I feel like chat GPT did most of the work as it generated the 10x10 matrix and the paths. We simply affected the aesthetics of the maze and simplified the generated solution." <br> "I felt that GPT contributed more. I only needed to give the prompt, and the GPT would produce an output without me having to think." |
| **Unsure** | Students did not clearly describe the proportion of their and ChatGPT's contribution, making is difficult to define who contributes more. Or their answers are irrelevant. | "Amazing! We had an awesome team leader who supervised and guided us throughout the process." <br> "Yes, I was specific on the parameters. And chat GPT gave me what I want." |

TABLE 2. CODING FRAMEWORK OF STUDENTS' INTERACTION EXPERIENCES WITH CHATGPT

| Interaction experiences | Descriptions | Examples |
|---|---|---|
| **Positive** | Students report positive experiences of their interaction with ChatGPT | "It was fun to see how AI-generated mazes and tried to provide solutions, although it was a new prompt for the AI to interact it." <br> "I enjoyed how fast the response was and how I was able to regenerate solutions endlessly." |
| **Mixed** | Students report negative experiences of their interaction with ChatGPT | "I feel like chat ChatGPT does a lot of things and I quite enjoyed the ease of usage. But I feel like the instructions to ChatGPT have to be made very clear, so it can sometimes be more of a hassle to instruct ChatGPT compared to outright doing it myself. SO perhaps for smaller less tedious questions I would choose to do it myself." |
| **Negative** | Students report both positive and negative experiences of their interaction with ChatGPT | "I find the ChatGPT always generates the one-glance-can-solve type of mazes instead of more challenging mazes. If I type instructions to increase the difficulty level, it will generate the unsolvable ones." <br> "I felt that the lack of support for maze design to be very frustrating as whatever we tried to generate ended up being unsolvable. This made it very hard to use the ideas that they provided." |
| **Unsure** | Students do not explictity describe their interaction experieces with ChatGPT. | "Generating multiple outputs from ChatGPT." <br> "I enjoyed working with my group mates." |

To address RQ3 concerning whether human-generative AI collaboration types, students' interaction experiences and sense of positive and negative agency predict their collaborative problem-solving, we first conducted Pearson correlation analyses of all the involved variables. Then, we ran a multiple linear regression analysis to predict collaborative problem-solving using all the other variables as independent variables. The analyses were conducted using R 4.1.2.

## IV. RESULTS

### A. Human-generative AI collaboration types

The pie chart in Fig. 2 presents the distribution of collaboration types between students and ChatGPT when working on problem-solving tasks. Nearly half of the students (44.30%) reported they led the collaboration with ChatGPT, indicating they predominantly directed the interaction, utilizing ChatGPT as a supportive tool to enhance their problem-solving efforts. A significant portion, 32.91%, reflects an 'even contribution' between students and ChatGPT, suggesting a synergistic approach where students and ChatGPT contributed comparably to the task, merging human creativity with AI's computational efficiency. Overall, the majority of students (77.21%) perceived they led or had even contributions when collaborating with ChatGPT. In 15.19% of the cases, 'AI leads' the collaboration. In this scenario, ChatGPT takes a more dominant role in the collaborative process, which may reflect a reliance on the AI's capabilities to guide the problem-solving process. In 7.59% of the cases, students did not adequately delineate the ratio of their own and ChatGPT's inputs, or their responses were not pertinent. The impacts of ChatGPT leading collaborative problem solving need to be further studied.

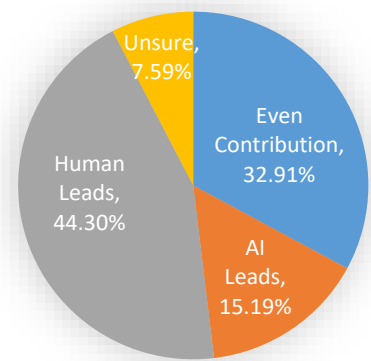

Fig. 2. Proportions of different human-generative AI collaboration types

### B. Students' interaction experiences with ChatGPT

As shown in Fig. 3, There were 27.85% of students who perceived their experiences as positive, implying that the interactions with ChatGPT were satisfactory for these individuals. About 39.24% of participants reported mixed experiences during their interaction with ChatGPT. This result suggests that these students encountered beneficial and challenging aspects when using ChatGPT, indicating a nuanced perception of the AI tool's capabilities and limitations. In total, more than two thirds of students perceived their interaction experiences with ChatGPT to be positive or mixed. There were 30.38% of students reported negative experiences, indicating they found the interaction with ChatGPT less satisfactory and the need to refine and adapt ChatGPT as well as instruct students how to better prompt for specific learning tasks. In 2.53% of the cases, students did not explicitly describe their interaction experiences and were coded as "unsure."

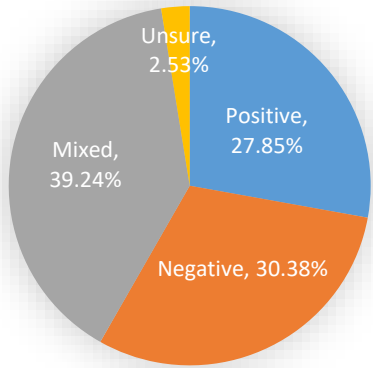

Fig. 3. Proportions of students' different interaction experiences with ChatGPT

### C. Factors predicting students' collaborative problem-solving

As Shown in Table 3, interaction experience has a significantly positive correlation with a sense of positive agency, indicating that more positive interaction experiences are related to a stronger sense of positive agency. Collaborative problem-solving is negatively correlated with a sense of negative agency, suggesting that a greater sense of negative agency is associated with students' perceived worse collaborative problem-solving. A sense of negative agency correlates negatively with a positive agency, with a statistically significant moderate correlation (r = -0.41**).

The multiple linear regression analysis aims to predict collaborative problem-solving using four predictors: human-generative AI collaboration types, interaction experience, sense of positive agency, and sense of negative agency. Sense of negative agency was a significant predictor, with higher levels of negative agency associated with lower collaborative problem-solving scores ($b$ = -0.34, $p$ = .006<.01). The model explained 10.8% of the variance in collaborative problem-solving, as indicated by an adjusted $R^2$ of .108, based on 72 observations.

TABLE 3. DESCRIPTIVE STATISTICS OF KEY VARIABLES

|  | ColType | InterExp | ColPS | SoPA | SoNA |
|---|---|---|---|---|---|
| **ColType** |  |  |  |  |  |
| **InterExp** | 0.02 |  |  |  |  |
| **ColPS** | -0.11 | 0.07 |  |  |  |
| **SoPA** | -0.13 | 0.26* | 0.22 |  |  |
| **SoNA** | -0.16 | -0.13 | -0.35** | -0.41*** |  |
| **M** | 2.32 | 1.92 | 4.12 | 3.81 | 2.62 |
| **SD** | 0.75 | 0.76 | 0.71 | 0.71 | 0.75 |

Note: ColType: human-generative AI collaboration types; InterExp: Interaction experience; ColPS: Collaborative problem solving; SoPA: Sense of positive agency; SoNA: Sense of negative agency

TABLE 4. REGRESSION RESULTS USING COLLABORATIVE INTERDISCIPLINARY PROBLEM-SOLVING AS THE DEPENDENT VARIABLE

|  | ColPS | | |
|---|---|---|---|
| *Predictors* | *Estimates* | *CI* | *p* |
| (Intercept) | 5.15 | 3.57 – 6.74 | **<0.001** |
| ColType | -0.16 | -0.38 – 0.06 | 0.162 |
| InterExp | 0.01 | -0.20 – 0.23 | 0.900 |
| SoPA | 0.05 | -0.21 – 0.31 | 0.701 |
| SoNA | -0.34 | -0.57 – -0.10 | **0.006** |
| Observations | 72 | | |
| $R^2$ / $R^2$ adjusted | 0.158 / 0.108 | | |

Note: ColType: human-generative AI collaboration types; InterExp: Interaction experience; ColPS: Collaborative problem solving; SoPA: Sense of positive agency; SoNA: Sense of negative agency

## V. DISCUSSION

Using the survey and reflection data of 79 undergraduate students, we examined three research questions regarding human-generative AI collaboration types and interaction experiences when students use ChatGPT for problem-solving tasks. Furthermore, we studied how collaboration types, interaction experiences, and sense of positive and negative agency correlated, and predicted students perceived collaborative problem-solving.

We found three human-generative AI collaboration types: even contribution, human leads, and AI leads, with even contribution and human leads being the majority (77.21%). This study has extended previous research [6] [11] by empirically examining the collaboration types between students and generative AI in the problem-solving context and quantifying the respective percentages of various collaboration types. Overall, most of the students perceived that they led problem solving, and at the same time, they were likely to leverage generative AI to augment their problem-solving skills. We also found about two-thirds of students had positive or mixed interaction experiences with ChatGPT. This result is consistent with previous research [23] [24] that suggests students generally found employing ChatGPT during their learning motivating, interesting, and helpful.

It should also be noted that 'AI leads' and 'negative experience' occupy unignorable proportions. 'AI leads' may suggest students' reliance on ChatGPT rather than taking leadership in the problem-solving process, which might negatively influence their learning and decision-making in the long term. Previous studies found the phenomenon of overreliance on AI, which may be related to having difficulty in evaluating AI's performance, trusting AI's recommendations when they should not, and changing their actions to align with AI's, and finally, and can result in unsatisfactory human and AI team performance [25] [26] [27]. ChatGPT might threaten students' critical thinking if they regarded it as the "ultimate epistemic authority" [28], and students could use ChatGPT to bypass reading and essay writing, which might result in superficial learning [29]. Further studies need to be conducted to examine who, when and why students rely on ChatGPT and what strategies can be taken to reduce their reliance on ChatGPT and mitigate its negative impacts. Efforts should also be made to improve technical (e.g., training and finetuning ChatGPT to improve its performance on specific tasks such as maze solving and math problems) and instructional design (e.g., helping students improve their prompt skills and understanding of how to use ChatGPT appropriately and ethically) regarding using ChatGPT for learning.

Surprisingly, only sense of negative agency negatively predicted collaborative problem-solving, while human-generative AI collaboration types and interaction experiences did not. These results suggest that how students collaborate with ChatGPT and their interaction experiences did not significantly influence their perceived collaborative problem-solving. However, whether students felt they could control their minds, actions and the environment when interacting with ChatGPT mattered to their perceived collaborative problem-solving. Therefore, it is critical to reduce students' sense of negative agency— a feeling that one's mind and the environment are not under their control. In our context, it is possible that the students relied on ChatGPT for the tasks and thus lost control over their own learning and group problem-solving tasks or because of the novelty of ChatGPT, some students spent too much time interacting with it. Similarly, Darvishi et al. [7] found that integrating AI into learning could influence students' sense of agency to take control of their own learning. They also advocated further studying the complex relationship between using AI and its impacts on agency. Attention should be paid to avoiding using AI for complete automation of the learning process and students' overreliance on AI. Instead, AI applications should empower students to develop agency, self-regulation skills and informed decision-making, for example, through personalized feedback and nudging systems [30] [31].

There are several limitations to this study. First, the participation rate was relatively low as only 8.74% of students who filled in the survey gave us consent to analyze their data, which may influence the sample's representativeness. Future research should replicate this study with larger sample sizes and higher participation rates to examine if the findings are reliable. Second, we used students' self-reported data to code their perceived human-generative AI collaboration types and prioritized students' perceptions of contributions (mainly referring to their writing) in the coding process. However, students might have different understandings and criteria of who led the work and what counted as even contributions, and their perceptions may differ from what happened. Therefore, future research should adopt more objective data (e.g., students' problem-solving log data and learning reports) to understand students' collaboration with AI. Similarly, the dependent variable, collaborative problem-solving, was measured using survey data. Future research should adopt more objective academic performance data to investigate the impacts of using ChatGPT differently on students' learning. With objective data, we will also research whether ChatGPT augments the problem-solving effectiveness and efficiency of student groups and whether it broadens or narrows down the problem-solving performance of groups.

Third, although the sample size is sufficient for the multiple linear regression and assumptions were checked and met, the limited sample size and available variables might not allow us to identify the most reliable and trustworthy predictors. Further research should replicate this study with a larger sample size and include more variables (e.g., students' engagement in problem solving and AI literacy).

Despite these limitations, the significance and implications of the study are two-fold. First, this study provides empirical evidence for different human-generative AI collaboration types and interaction experiences in the problem-solving context. This result suggests the need for future research and practice to facilitate students to appropriately use ChatGPT and other generative AI tools and finetune the tools to improve their understanding of and responses to specific tasks. Second, this study shows the negative impacts of a sense of negative agency on students' perceived collaborative problem-solving, which calls for further research to understand why students perceive negative agency and how to reduce this perception. When using ChatGPT for learning, educators and researchers should not only pay attention to curriculum design and the effectiveness of the tool but also to students' psychological factors, such as sense of agency and self-regulation.

## VI. Conclusion

This study expanded previous research by (1) providing empirical evidence of the existence of various collaboration types between human and generative AI and their frequency; (2) extending the tools from AI to generative AI, which tends to be more versatile; and (3) including sense of positive agency and sense of negative agency in the analyses to examine how they are correlated with collaboration types, interaction experiences, and collaborative problem-solving. We found that 77.21% of students perceived their collaboration with ChatGPT as human-leads or even contributions, whereas 15.19% as AI-leads. Sense of negative agency negatively predicted Collaborative problem solving. These results suggest the phenomenon of students losing control over and relying on ChatGPT during problem-solving and its negative impacts. Future research should further study how to support students to use generative AI appropriately and ethically without harming their agency and problem-solving skills.

## Acknowledgment

We are indebted to the participants who made this study possible. This study is supported by NTU EdeX EdeX Teaching and Learning Grant (No. 1/22 ZG).

## References

1. van den Bosch, K., et al. *Six challenges for human-AI Co-learning*. in *Adaptive Instructional Systems: First International Conference, AIS 2019, Held as Part of the 21st HCI International Conference, HCII 2019, Orlando, FL, USA, July 26–31, 2019, Proceedings 21*. 2019. Springer.
2. Dellermann, D., et al., *Hybrid intelligence.* Business & Information Systems Engineering, 2019. **61**: p. 637-643.
3. Dellermann, D., et al., *The future of human-AI collaboration: a taxonomy of design knowledge*

*for hybrid intelligence systems.* arXiv preprint arXiv:2105.03354, 2021.
4. Graesser, A.C., et al., *Collaboration in the 21st century: The theory, assessment, and teaching of collaborative problem solving.* 2020, Elsevier. p. 106134.
5. Seeber, I., et al., *Machines as teammates: a collaboration research agenda.* 2018.
6. Cheng, X. and S. Zhang, *Tool, Teammate, Superintelligence: Identification of ChatGPT-Enabled Collaboration Patterns and their Benefits and Risks in Mutual Learning.* 2024.
7. Darvishi, A., et al., *Impact of AI assistance on student agency.* Computers & Education, 2023: p. 104967.
8. Fügener, A., et al., *Cognitive challenges in human-AI collaboration: Investigating the path towards productive delegation.* Forthcoming, Information Systems Research, 2019.
9. Memmert, L. and E. Bittner, *Complex Problem Solving through Human-AI Collaboration: Literature Review on Research Contexts.* 2022.
10. Ouyang, F. and P. Jiao, *Artificial intelligence in education: The three paradigms.* Computers and Education: Artificial Intelligence, 2021. **2**: p. 100020.
11. Liu, D., Bridgeman, Adam. *ChatGPT is old news: How do we assess in the age of AI writing co-pilots?* 2023 [cited 2023 21 August].
12. Steyvers, M. and A. Kumar, *Three Challenges for AI-Assisted Decision-Making.* 2022.
13. Celik, I., et al., *The promises and challenges of artificial intelligence for teachers: A systematic review of research.* TechTrends, 2022. **66**(4): p. 616-630.
14. Fügener, A., et al., *Will humans-in-the-loop become borgs? Merits and pitfalls of working with AI.* Management Information Systems Quarterly (MISQ)-Vol, 2021. **45**.
15. Benbya, H., S. Pachidi, and S. Jarvenpaa, *Special issue editorial: Artificial intelligence in organizations: Implications for information systems research.* Journal of the Association for Information Systems, 2021. **22**(2): p. 10.
16. Synofzik, M., G. Vosgerau, and M. Voss, *The experience of agency: an interplay between prediction and postdiction.* Frontiers in psychology, 2013. **4**: p. 127.
17. Vaughn, M., *What is student agency and why is it needed now more than ever?* Theory Into Practice, 2020. **59**(2): p. 109-118.
18. Code, J., *Agency for learning: Intention, motivation, self-efficacy and self-regulation.* Frontiers in Genetics, 2020. **5**: p. 19.
19. Rahman, M.M., *21st century skill'problem solving': Defining the concept.* Rahman, MM (2019). 21st Century Skill "Problem Solving": Defining the Concept. Asian Journal of Interdisciplinary Research, 2019. **2**(1): p. 64-74.
20. Cole, M.L., J.D. Cox, and J.M. Stavros, *SOAR as a mediator of the relationship between emotional intelligence and collaboration among professionals working in teams: Implications for entrepreneurial teams.* Sage Open, 2018. **8**(2): p. 2158244018779109.
21. Tavakol, M. and R. Dennick, *Making sense of Cronbach's alpha.* International journal of medical education, 2011. **2**: p. 53.
22. Tapal, A., et al., *The sense of agency scale: A measure of consciously perceived control over one's mind, body, and the immediate environment.* Frontiers in psychology, 2017. **8**: p. 1552.
23. Ali, J.K.M., et al., *Impact of ChatGPT on learning motivation: teachers and students' voices.* Journal of English Studies in Arabia Felix, 2023. **2**(1): p. 41-49.
24. Shoufan, A., *Exploring Students' Perceptions of CHATGPT: Thematic Analysis and Follow-Up Survey.* IEEE Access, 2023.
25. Buçinca, Z., M.B. Malaya, and K.Z. Gajos, *To trust or to think: cognitive forcing functions can reduce overreliance on AI in AI-assisted decision-making.* Proceedings of the ACM on Human-Computer Interaction, 2021. **5**(CSCW1): p. 1-21.
26. Bansal, G., et al. *Is the most accurate ai the best teammate? optimizing ai for teamwork.* in *Proceedings of the AAAI Conference on Artificial Intelligence*. 2021.
27. Passi, S. and M. Vorvoreanu, *Overreliance on AI Literature Review.* Microsoft Research, 2022.
28. Cooper, G., *Examining science education in chatgpt: An exploratory study of generative artificial intelligence.* Journal of Science Education and Technology, 2023. **32**(3): p. 444-452.
29. Dwivedi, Y.K., et al., *"So what if ChatGPT wrote it?" Multidisciplinary perspectives on opportunities, challenges and implications of generative conversational AI for research, practice and policy.* International Journal of Information Management, 2023. **71**: p. 102642.
30. Wambsganss, T., A. Janson, and J.M. Leimeister, *Enhancing argumentative writing with automated feedback and social comparison nudging.* Computers & Education, 2022. **191**: p. 104644.
31. Afzaal, M., et al., *Explainable AI for data-driven feedback and intelligent action recommendations to support students self-regulation.* Frontiers in Artificial Intelligence, 2021. **4**: p. 723447.